\documentclass[conference]{IEEEtran}
\ifCLASSINFOpdf
 \usepackage[pdftex]{graphicx}
\else
\fi

\usepackage{caption}
\usepackage{subcaption}

\usepackage{amsmath, amsthm, amssymb}
\newtheorem{thm}{Theorem}
\newtheorem{lem}[thm]{Lemma}
\newtheorem{cor}[thm]{Corollary}
\newtheorem{prop}[thm]{Proposition}

\begin{document}
\title{LDPC Codes for Partial-Erasure Channels in Multi-Level Memories}

\author{\IEEEauthorblockN{Rami Cohen and Yuval Cassuto}
\IEEEauthorblockA{Department of Electrical Engineering\\
Technion - Israel Institute of Technology\\
Technion City, Haifa 3200000, Israel\\
Email: rc@tx.technion.ac.il, ycassuto@ee.technion.ac.il}}


\maketitle

\begin{abstract}
%\boldmath
In this paper, we develop a new channel model, which we name the $q$-ary partial erasure channel (QPEC). QPEC has a $q$-ary input, and its output is either one symbol or a set of $M$ possible values. This channel mimics situations when current/voltage levels in measurement channels are only partially known, due to high read rates or imperfect current/voltage sensing. Our investigation is concentrated on the performance of low-density parity-check (LDPC) codes when used over this channel, due to their low decoding complexity with iterative-decoding algorithms. We give the density evolution equations of this channel, and develop its decoding-threshold analysis. Part of the analysis shows that finding the exact decoding threshold efficiently lies upon a solution to an open problem in additive combinatorics. For this part we give bounds and approximations.  
\end{abstract}
\IEEEpeerreviewmaketitle

\section{Introduction}
The advent of non-volatile memories (NVMs) with many levels per cell holds a great promise for increased storage capacity. At the same time, it proves extremely challenging to write and read many-level cells at both high precision and high speeds. As a result, coding is employed to improve the tradeoff between data reliability and access speed (see e.g.~\cite{YC_ALM:10}). Natural candidates to improve the reliability of NVMs are low-density parity-check (LDPC) codes~\cite{Gallager1}, which offer low complexity of implementation and good performance under iterative decoding \cite{mct}. Recent work on the employment of LDPC codes in NVMs, such as~\cite{Wessel}, focused on the additive white Gaussian noise (AWGN) channel. 

In addition to the AWGN and other classical channels, NVMs motivate coding for a diversity of new channels with rich features. Our work here is motivated by a class of channels we call {\em measurement channels}, which encompass a variety of equivocations introduced to the information by an imperfect read process. This imperfection of the read process comes from either physical limitations or speed constraints. In particular, the channel model we study here -- the {\em $q$-ary partial erasure channel} (QPEC) -- comes from a read process that occasionally fails to read the information at its entirety, and provides as decoder inputs $q$-ary symbols that are {\em partially} erased.      

Theoretically speaking, the QPEC is an extension of the $q$-ary erasure channel (QEC)~\cite{bennatan}, where instead of erasing a full channel symbol, the channel returns a set of $M\leq q$ symbols that contains the correct stored symbol and $M-1$ other symbols. Our results on the QPEC include calculating its capacity in Section \ref{QPEC_DEF}, a message-passing decoder in Section \ref{message_passing}, and analysis and approximation models for its density-evolution formulation in Sections \ref{de_analysis} and \ref{pm_approx}. 
\section{QPEC: Q-ary Partial Erasure Channel}
\label{QPEC_DEF}
\subsection{Channel model}
\label{channel_model}
The {\it Q-ary Partial Erasure Channel} (QPEC) is defined as follows. Let $X$ be the transmitted symbol taken from the alphabet $\mathcal{X} = \left\{ {0,1,...,q - 1} \right\}$. Let $Y$ be the received symbol with the output alphabet $\mathcal{Y} = \left\{ {\mathcal{X}\bigcup\limits_{x = 0}^{q - 1} {\left\{ {?_x^{\left( i \right)}} \right\}_{i = 1}^{{i_{\max }}}} } \right\}$, where each super-symbol ${?_x^{\left(i \right)}}$ (for $x=0,1,...,q-1$) consists of a set of size $M$ that contains the symbol $x$ and $M-1$ other symbols, taken from $\mathcal{X}\backslash \left\{ x \right\}$. Let's denote by $\ell \left( {n,k} \right)$ the binomial coefficient $\left( {\begin{array}{*{20}{c}}
n\\
k
\end{array}} \right)$. Clearly, ${i_{\max }} = \ell \left( {q - 1,M - 1} \right)$.  

The transition probabilities governing the QPEC are as follows:
\begin{equation}
\label{tran_matrix}
\Pr \left( {\left. {Y = y} \right|X = x} \right) = \left\{ {\begin{array}{*{20}{c}}
{1 - \varepsilon ,}&{y = x}\\
{\varepsilon /{i_{\max }},}&{y = ?_x^{\left(i \right)}}
\end{array}} \right.
\end{equation}
for $i=1,2,...,i_{\rm{max}}$, where $0 \le \varepsilon  \le 1$ is the (partial) erasure probability. That is, the output of the channel can be either a {\it symbol}, with probability $1-\varepsilon$ (corresponding to a non-erasure event), or a {\it set of $M$ symbols}, with probability $\varepsilon$ (corresponding to a partial erasure event). As an example, assume that $q=4, M=2$, and the symbol $0$ was transmitted. Then we have $?_x^{\left( 1 \right)} = \left\{ {0,1} \right\},?_x^{\left( 2 \right)} = \left\{ {0,2} \right\}$ and $?_x^{\left( 3 \right)} = \left\{ {0,3} \right\}$, where each is received with probability $\varepsilon/3$ and $0$ is received with probability $\varepsilon$.
%Note that  can be generalized to a channel with multiple $\left( {{M_i},{\varepsilon _i}} \right)$ pairs. In this case, we will have an output of size $M_i$ with probability $\varepsilon_i$, and a set of size $1$ (one symbol) with probability $1 - \sum\limits_i {{\varepsilon _i}} $. Most of our results in this paper can be generalized to this case. 

Note that for $M=q$ we get the $q$-ary erasure channel (QEC), the common generalization of the BEC to $q>2$. In our analysis, we will use the arithmetic of the finite field GF($q$), such that $q$ will be a prime or a prime power, and the symbol alphabet will be assumed to be the elements of GF($q$).

\subsection{Capacity}
Denote $p_k=\Pr(X = k)$, for $k = 0,1,...,q-1$, to be the
input distribution to the channel. According to the definition of the channel capacity $C$:
\begin{equation}
\label{cap_def}
C = {\max _{\left\{ {{p_k}} \right\}_{k = 0}^{q - 1}}}I\left( {X;Y} \right) = {\max _{\left\{ {{p_k}} \right\}_{k = 0}^{q - 1}}}\left( {H\left( Y \right) - H\left( {\left. Y \right|X} \right)} \right),
\end{equation}
where $I\left( {X;Y} \right)$ is the mutual information between the input $X$ and the output $Y$, and $H\left( Y \right)$, ${H\left( {\left. Y \right|X} \right)}$ are the entropy of $Y$ and the conditional entropy of $Y$ given $X$, respectively. The conditional entropy ${H\left( {\left. Y \right|X} \right)}$ can be calculated using \eqref{tran_matrix}:
\begin{equation}
\label{cond_entropy}
H\left( {\left. Y \right|X} \right) =  - \left( {1 - \varepsilon } \right)\log \left( {1 - \varepsilon } \right) - \varepsilon \log \left( {\varepsilon /{i_{\max }}} \right)
\end{equation}
which is independent of input distribution (as expected), implying that it is sufficient to maximize the entropy ${H\left( Y \right)}$. Similarly to the case of the BEC, ${H\left( Y \right)}$ is maximized under the uniform distribution of the input.
\begin{thm}
%\begin{theorem}
\label{capacity_achieving}
\emph{(Capacity achieving input distribution for the QPEC)} Assume a QPEC channel with an input probability distribution ${\left\{ {{p_k}} \right\}_{k = 0}^{q - 1}}$. Then the capacity is achieved for the uniform distribution of the input, and we have:
\begin{equation}
C\left( {{\text{QPEC}}} \right) = 1 - \varepsilon {\log _q}M
\end{equation}
measured in $q$-ary symbols per channel use.
\end{thm}
%\end{theorem}
\begin{proof}
Denote:
\[A = 1 - \varepsilon ,B = \frac{\varepsilon }{{\ell \left( {q - 1,M - 1} \right)}},I = \ell \left( {q,M} \right)\]

In addition, define the sets $S_i$, for $i=1,2,...,I$, where each set contains $M$ distinct elements taken from the set $\left\{ {0,1,...,q - 1} \right\}$, such that $S_i \ne S_j$ for $i \ne j$. Since $H(Y)$ is a function of the input distribution only (when $q,M,\varepsilon$ are given), we are able to define $f\left( {\left\{ {{p_k}} \right\}} \right) \buildrel \Delta \over = H\left( Y \right)$. We now have:
\begin{equation}
\begin{array}{*{20}{l}}
{f\left( {\left\{ {{p_k}} \right\}_{k = 0}^{q - 1}} \right) =  - \sum\limits_{k = 0}^{q - 1} {A{p_k}\log \left( {A{p_k}} \right)} }\\
{ - \sum\limits_{i = 1}^I {\left( {B\sum\limits_{j \in {S_i}} {{p_j}} } \right)\log \left( {B\sum\limits_{j \in {S_i}} {{p_j}} } \right)} }
\end{array}
\end{equation}
so that ${\left\{ {{p_k}} \right\}_{k = 0}^{q - 1}}$ can be found by solving the following maximization problem:
\begin{equation}
{\max _{\left\{ {{p_k}} \right\}_{k = 0}^{q - 1}}}f\left( {\left\{ {{p_k}} \right\}_{k = 0}^{q - 1}} \right),\hspace{7pt} {\rm{s}}.{\rm{t}}.\sum\limits_{k = 0}^{q - 1} {{p_k}}  = 1
\end{equation}
Using the method of Lagrange multipliers, we get the following system of equations:
\begin{equation}
\frac{{\partial f}}{{\partial {p_k}}} + \lambda  = 0,\hspace{7pt} \sum\limits_{k = 0}^{q - 1} {{p_k}}  = 1
\end{equation}
where $\lambda$ is the Lagrange multiplier. The equations translate into:
\begin{equation}
\begin{array}{l}
 - A\log \left( {{p_k}} \right) - Aq - \sum\limits_{{S_i}:k \in {S_i}} {B\log \left( {B\sum\limits_{j \in {S_i}} {{p_j}} } \right)}  \\
 - B\left( {I - 1} \right) + \lambda  = 0,\hspace{7pt} \sum\limits_{k = 0}^{q - 1} {{p_k}}  = 1
\end{array}
\end{equation}
meaning that:
$$ - A\log \left( {{p_k}} \right) - Aq - \sum\limits_{{S_i}:k \in {S_i}} {B\log \left( {B\sum\limits_{j \in {S_i}} {{p_j}} } \right) - B\left( {I - 1} \right)} $$
are equal for all $k$. This is satisfied when $p_k=1/q$ for $k=0,1,...,q-1$. In addition, $I\left( {X;Y} \right)$ is a concave function of $p_k$ once $\Pr \left( {\left. {Y = y} \right|X = x} \right)$ is given, and therefore $p_k=1/q$ leads to the global maximum of $I\left( {X;Y} \right)$, that is, to the capacity.
\end{proof}

%\begin{theorem}
%\label{QPEC_cap}
%\emph{(Capacity of the QPEC)} Assume a QPEC with parameters $\left( {q,M,\varepsilon } \right)$. Then: 

%\end{theorem}
%\begin{proof}
%By using Lemma \ref{capacity_achieving} and its notations, and the conditional entropy \eqref{cond_entropy}, we get for $p_k=1/q$:
%\begin{equation}
%\begin{array}{*{20}{l}}
%{C = f\left( {\left\{ {{p_k}} \right\}_{k = 0}^{q - 1}} \right) - H\left( {Y|X} \right) = \left( {1 - \varepsilon } \right){{\log }_q}q + \varepsilon {{\log }_q}B}\\
%{ - \varepsilon {{\log }_q}\frac{{MB}}{q} = 1 - \varepsilon {{\log }_q}M}
%\end{array}
%\end{equation}
Note that the capacity $C$ for the QPEC is in agreement with the capacity of the QEC ($M=q$) and  in particular with the capacity of the BEC ($M=q=2$).

\section{Message Passing Algorithm for the QPEC}
\label{message_passing}

A GF($q$) LDPC $\left[ {n,k} \right]$ code is defined in a similar way to its binary counterpart, by a sparse parity-check matrix, or equivalently by a Tanner graph \cite{tanner}. This graph is bipartite, with $n$ variable (left) nodes, which correspond to symbols of the codeword, and $n-k$ check (right) nodes, which correspond to parity check equations. 
The codeword symbols and the labels on the edges of the graph are taken from GF($q$). For ease of presentation we will concentrate here on \textit{regular} LDPC codes, having a constant check node degree $d_c$ and a constant variable node degree $d_v$.

In the graph, each check node $c_j$ is connected, by edges, to variable nodes $v_i, i \in N(j)$, where $N(j)$ denotes the set of nodes adjacent to node $i$. The parity check equation induced by $c_j$ is satisfied when $\sum\limits_{i \in N\left( j \right)} {{h_{ij}}{v_i}}  = 0$, where $h_{ij}$ is the label on the edge connecting variable node $i$ to check node $j$.

The following decoder for $q$-ary LDPC codes over the QPEC is a variation of the standard message passing/belief propagation algorithm over a Tanner graph to match the partial information exchanged in decoding. For this decoder, the beliefs exchanged in the decoding process are {\it sets of symbols}, rather than probabilities. We have two types of messages: \textit{check to variable} (CTV) messages, and \textit{variable to check} (VTC) messages, denoted by $c_{j \to i}$ and $v_{i \to j}$, respectively.
%For a variable node $i$ and a check node $j$, we denote by ${v_{i \to j}}$ a VTC message and by ${c_{j \to i}}$ a CTV message. Each outgoing message from a variable (check) node to a check (variable) node depends on all its incoming message, except for the incoming message originated from the target node.

\begin{figure}
        \centering
        \begin{subfigure}[b]{0.15\textwidth}
                \includegraphics[width=\textwidth]{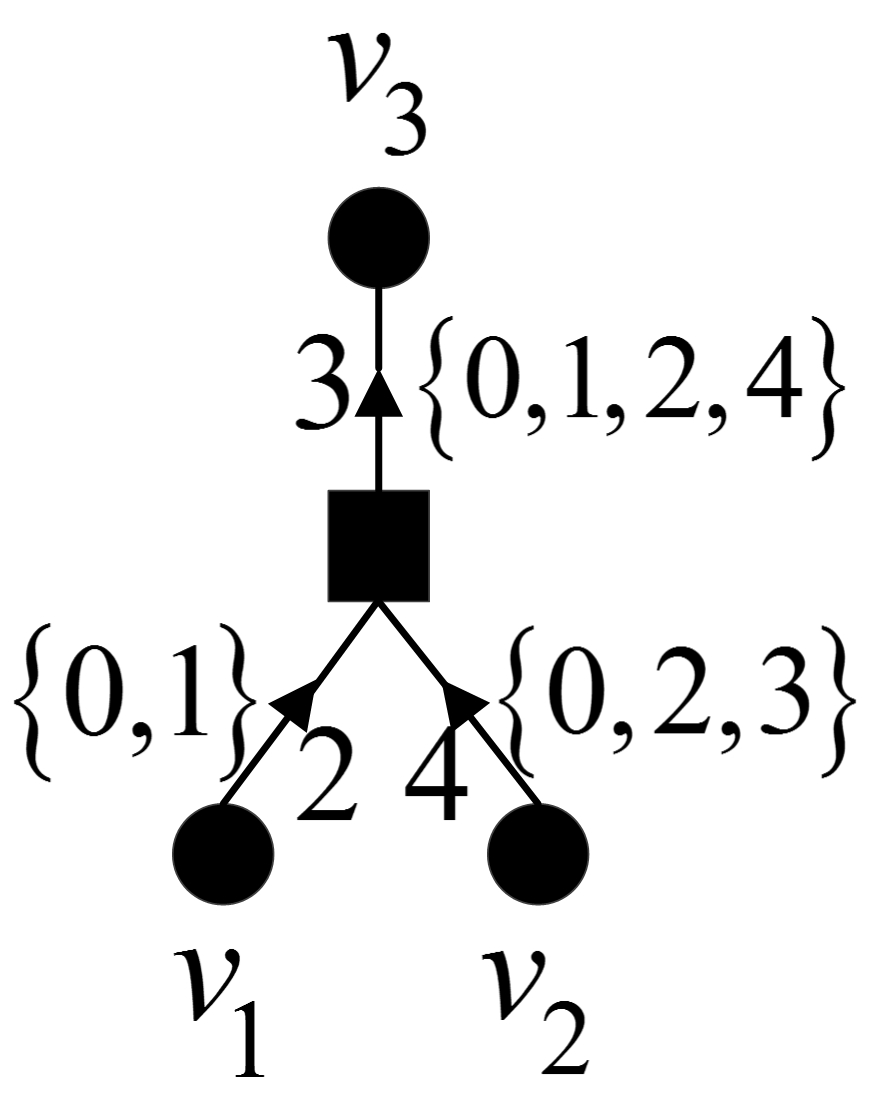}
                \caption{CTV message}
               \label{fig_CTV}
        \end{subfigure}%
        ~ %add desired spacing between images, e. g. ~, \quad, \qquad etc.
        \begin{subfigure}[b]{0.15\textwidth}
                \includegraphics[width=\textwidth]{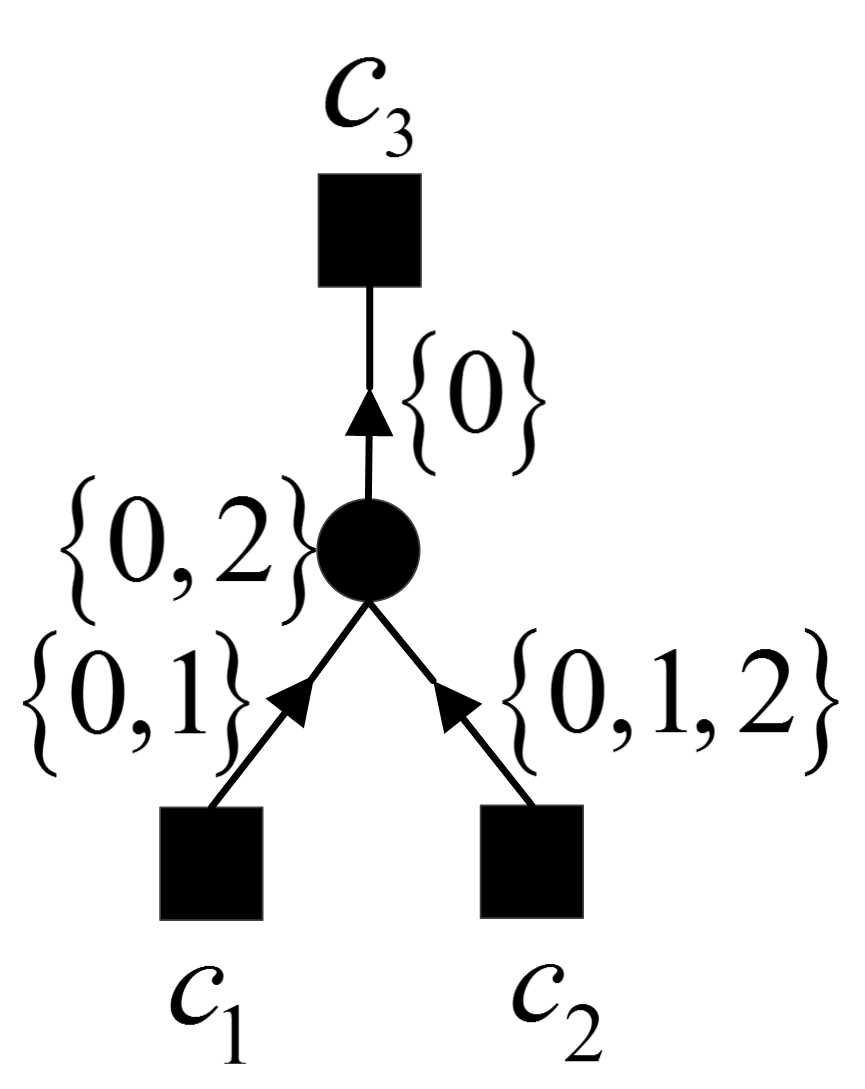}
                \caption{VTC message}
                \label{fig_VTC}
        \end{subfigure}

        \caption{Message passing: examples (GF($5$))}
\label{msg_ex}
\end{figure}

At iteration $l=0$, channel information is sent from variable to check nodes: erased nodes send sets of size $M$, and non-erased ones send sets of size $1$ (containing the correct symbol). In the next iterations, we have the following messages:

\begin{enumerate}
\item {\bf Check to variable (CTV)}. 

%The outgoing message ${c_{j \to i}}$ from check node $j$ to variable node $i$ is a set consisting of all possible symbols for the target variable node, such that the . 
First, we define for each ${i' \in N\left( j \right)\backslash i}$ the following set:
\begin{equation}
\label{X_i_def}
X_{i'}^{(l)} = \left\{ { - \frac{{{h_{i'j}} \cdot {x_{i'}}}}{{{h_{ij}}}}:{x_{i'}} \in v_{i' \to j}^{\left( {l - 1} \right)}} \right\}, \hspace{7pt} l \ge 1.
\end{equation}

Then we have:
\begin{equation}
\label{CTV}
\begin{array}{l}
c_{j \to i}^{\left( l \right)} = \sum\limits_{i' \in N\left( j \right)\backslash i} {X_{i'}^{\left( l \right)}} \\
 \buildrel \Delta \over = \left\{ {\sum\limits_{i' \in N\left( j \right)\backslash i} {{a_{i'}}:{a_{i'}} \in X_{i'}^{\left( l \right)}} } \right\},\hspace{7pt} l \ge 1
\end{array}
\end{equation}

where the calculations are carried over GF($q$). In words, the message \eqref{CTV} consists of all possible assignments of the variable node $i$,  such that the parity equation involving the variable nodes in the set $N\left( j \right)$ is satisfied. 

An example for a CTV message is given in Figure \ref{fig_CTV}. In this example (over GF($5$)), the variable nodes $v_1, v_2$ and $v_3$ are connected to the same check node, with edges having the labels $2, 4$ and $3$. $v_1$ is known to be either $0$ or $1$, and $v_2$ is known to be either $0$, $2$ or $3$. Therefore, the outgoing message is consisted of all possible outcomes of the expression $\left( {2{v_1} + 4{v_2}} \right)/\left( { - 3} \right)$.
%
%\begin{figure}[!t]
%\centering
%\includegraphics[scale=0.6]{CTV.jpg}
%\caption{Example for a check to variable (CTV) message.}
%\label{fig_CTV}
%\end{figure}

\item {\bf Variable to check (VTC)}. 
%The outgoing message is ${v_{i \to j}}$ from variable node $i$ to check node $j$ is the intersection set of the incoming messages from all neighbouring check nodes, except $j$:
\begin{equation}
\label{VTC}
v_{i \to j}^{\left( l \right)} = v_{i \to j}^{\left( 0 \right)}\bigcap {\left( {\bigcap\limits_{j' \in N\left( i \right)\backslash j} {c_{j' \to i}^{\left( l \right)}} } \right)}, \hspace{10pt} l \ge 1
\end{equation}
where ${v_{i \to j}^{\left( 0 \right)}}$ is the output from the channel information for variable node $i$, which is passed at iteration $0$. The resulting message is simply the intersection of the incoming messages and the channel information. An example for a VTC message is given in Figure \ref{fig_VTC}.

%\begin{figure}[!t]
%\centering
%\includegraphics[scale=0.6]{VTC.jpg}
%\caption{Example for a variable to check (VTC) message (the labels were omitted). The set at the variable node is the channel information.}
%\label{fig_VTC}
%\end{figure}
\end{enumerate}

In practice, the decoder stops after a finite number of iterations. The decoding is declared successful if the size of all VTC messages is $1$.%, as we show in the following lemma.

%\begin{lemma}
%\emph{(Sufficient condition for correct decoding)}
%Assume a ($q,M,\varepsilon$)-QPEC channel. If at iteration $l=l_0$ $\left| {v_{i \to j}^{\left( {{l_0}} \right)}} \right| = 1$ for all variable nodes, then the transmitted codeword is decoded correctly with probability $1$ at iteration $l_0$.
%\label{suff_decod}
%\end{lemma}
%\begin{proof}
%By definition, the transmitted (correct) symbol for each variable node appears in its initial set. This is preserved under the union operation done in check nodes, since the assignment of symbols to their outgoing sets must contain the correct symbol (according to the parity check equation). This property is preserved as well under the intersection operation done in the case of variable nodes. Therefore, once we are left with sets of size $1$ at each variable node, it is guaranteed that the codeword is decoded correctly.
%\qed
%\end{proof}

\section{Decoding Analysis through Density Evolution}
\label{de_analysis}

The density evolution method proposed in \cite{RU} is an analytical tool for evaluating the asymptotic performance of LDPC codes under message-passing decoding. Note that in our case the \textit{all-zero codeword assumption} \cite{RU} holds, since the noise is independent of the transmitted codeword. %See [xx] for details.

The key idea we use for analyzing the densities is to track the probability distribution on the {\it sizes} of the messages, leading to just $q$ entries in the distribution, instead of $2^q-1$. This approach is a more natural one in our case, since a decoding failure may occur when a VTC message has size larger than $1$, independent of the exact content of the message.

\subsection{Density-evolution equations}
\label{de_equations}

In this part, we present the density evolution equations corresponding to BP decoding for the QPEC, assuming that the LDPC graph was drawn at random. In the following, $\mathbf{w}^{(l)}$ is a probability vector, where $w_m^{\left( l \right)}$ ($m=1,2,...,q$) denotes the probability that a CTV message at iteration $l$ is of size $m$. The probability vector ${\bf{z}}^{\left( l \right)}$ is defined for VTC messages in a similar manner. 

%The following density-evolution equations are based on the probabilities for outgoing messages of size $m$. For CTV messages, we denote this probability by $P_m$, and for VTC messages it is denoted by $Q_m$. We finally get:

The following density-evolution equations are based on the following idea. For each possible set of sizes of incoming messages, its probability is calculated by multiplying the probability of each incoming message size. This probability is then multiplied by the probability that the outgoing message will be of size $m$, given the sizes of the incoming messages. We get:

\begin{enumerate}
\item {\bf CTV messages:}
\begin{equation}
\label{Pm1}
w_m^{\left( l \right)} = \sum\limits_{\left\{ {{S_j}} \right\}_{j = 1}^{{d_c} - 1}} {\left( {\prod\limits_{j = 1}^{{d_c} - 1} {z_{\left| {{S_j}} \right|}^{\left( {l - 1} \right)}} } \right) \cdot {P_m}\left( {\left\{ {\left| {{S_j}} \right|} \right\}_{j = 1}^{{d_c} - 1}} \right)} 
\end{equation}
%w_m^{\left( l \right)} = \sum\limits_{\scriptstyle\left\{ {{S_j}} \right\}_{j = 1}^{{d_c} - 1}:\hfill\atop
%{\scriptstyle{S_j} \subseteq {\rm{GF}}\left( q \right),\hfill\atop
%\scriptstyle\left| {{S_j}} \right| \le M\hfill}} {\left( {\prod\limits_{j = 1}^{{d_c} - 1} {z_{\left| {{S_j}} \right|}^{\left( {l - 1} \right)}} } \right) \cdot {P_m}\left( {\left\{ {\left| {{S_j}} \right|} \right\}_{j = 1}^{{d_c} - 1}} \right)} 
such that ${S_j} \subseteq {\rm{GF}}\left( q \right)$ and $\left| {{S_j}} \right| \le M$. $P_m$ denotes the probability that a CTV message is of size $m$, given the sizes of the incoming VTC messages, ${\left\{ {\left| {{S_j}} \right|} \right\}_{j = 1}^{{d_c} - 1}}$.

\item {\bf VTC messages:}

\label{VTC_DE}

\begin{equation}
\label{Qm_DE}
\begin{array}{l}
z_m^{\left( l \right)} = \delta \left[ {m - 1} \right] \cdot \left( {1 - \varepsilon } \right)\\
 + \varepsilon \sum\limits_{\left\{ {{S_j}} \right\}_{j = 1}^{{d_v} - 1}} {\left( {\prod\limits_{j = 1}^{{d_v} - 1} {w_{\left| {{S_j}} \right|}^{\left( l \right)}} } \right) \cdot {Q_m}\left( {\left\{ {\left\{ {\left| {{S_j}} \right|} \right\}_{j = 1}^{{d_v} - 1},M} \right\}} \right)} 
\end{array}
\end{equation}

such that ${S_j} \subseteq {\rm{GF}}\left( q \right)$ and $\left| {{S_j}} \right| \le q$. $\delta \left[ m \right]$ denotes the discrete Dirac delta function. $Q_m$ denotes the probability that a VTC message is of size $m$, given the sizes of the incoming CTV messages, ${\left\{ {\left| {{S_j}} \right|} \right\}_{j = 1}^{{d_v} - 1}}$, and the size $M$ of the partially-erased variable node.
\end{enumerate}

Finding the exact $P_m$ as a function of the incoming message sizes is a hard problem, as we will see in Section \ref{pm_bounds}, where several bounds over $P_m$ will be given. We also give two approximation models for $P_m$ in Section \ref{pm_approx}. On the other hand, an exact formula for $Q_m$ will be provided in Section \ref{PDVTC}. 

Note that regardless of the exact behaviour of $P_m$, the density-evolution equation of the BEC \cite{mct} (hence QEC) are equivalent to Equations \eqref{Pm1} and \eqref{Qm_DE} when $M=q$. 
For showing this equivalence, we consider irregular LDPC codes, by defining the following two polynomials \cite{mct}:
\begin{equation}
\label{lambda_eq}
{\lambda \left( x \right) = \sum\limits_{i = 2}^{{d_v}} {{\lambda _i}{x^{i - 1}}} }\\
\end{equation}
\begin{equation}
\label{rho_eq}
{\rho \left( x \right) = \sum\limits_{i = 2}^{{d_c}} {{\rho _i}{x^{i - 1}}} }
\end{equation}
where for each $i$, a fraction $\lambda_i$ ($\rho_i$) of the edges is connected to variable (check) nodes of degree $i$. Note that $d_v$ ($d_c$) denotes now the \textit{maximal} degree of a variable (check) node.

When $M=q$, the QPEC BP messages are of size $1$ or size $q$ only. Therefore, ${{w}}_2^{\left( l \right)},{{w}}_3^{\left( l \right)},...,{{w}}_{q - 1}^{\left( l \right)} = 0$ and ${{z}}_2^{\left( l \right)},{{z}}_3^{\left( l \right)},...,{{z}}_{q - 1}^{\left( l \right)} = 0$ for all $l$. Extending Equations \eqref{Pm1} and \eqref{Qm_DE} to the irregular case, we have:
\begin{equation}
\label{Mq_case}
  {{w}}_1^{\left( l \right)} = \sum\limits_{i = 2}^{{d_c}} {{\rho _i}} {\left( {{{z}}_1^{\left( {l - 1} \right)}} \right)^{i - 1}} \hfill 
\end{equation}
\begin{equation}    
\label{Mq_case1}
  {{z}}_q^{\left( l \right)} = \varepsilon \sum\limits_{i = 2}^{{d_v}} {{\lambda _i}} {\left( {{{w}}_q^{\left( l \right)}} \right)^{i - 1}} \hfill \\ 
\end{equation}
Plugging Equation \eqref{Mq_case} into Equation \eqref{Mq_case1}, we get:
\begin{equation}
\begin{array}{l}
z_q^{\left( l \right)} = \varepsilon \sum\limits_{i = 2}^{{d_v}} {{\lambda _i}} {\left( {1 - w_1^{\left( l \right)}} \right)^{i - 1}}\\
 = \varepsilon \sum\limits_{i = 2}^{{d_v}} {{\lambda _i}} {\left( {1 - \sum\limits_{j = 2}^{{d_c}} {{\rho _j}} {{\left( {z_1^{\left( {l - 1} \right)}} \right)}^{j - 1}}} \right)^{i - 1}}\\
 = \varepsilon \sum\limits_{i = 2}^{{d_v}} {{\lambda _i}} {\left( {1 - \sum\limits_{j = 2}^{{d_c}} {{\rho _j}} {{\left( {1 - z_q^{\left( {l - 1} \right)}} \right)}^{j - 1}}} \right)^{i - 1}},
\end{array}
\end{equation}
leading to the well known recurrence relation for the BEC (QEC with $M=q=2$) (as derived in \cite{Luby97}, \cite{Luby98}), which holds for the QEC as well:
\begin{equation}
P_e^{\left( l \right)} = \varepsilon \lambda \left( {1 - \rho \left( {1 - P_e^{\left( {l - 1} \right)}} \right)} \right), \hspace{10pt} l \ge 1
\end{equation}
where $P_e^{(l)}$ is the decoding failure probability at iteration $l$.

\subsection{Equivalent formulation for $P_m$ and bounds}
\label{pm_bounds}

Assume that we have $K$ subsets of GF($q$), $\left\{ {{S_j}} \right\}_{j = 1}^K$. Their \textit{sumset}, denoted $\sum\limits_{j = 1}^K {{S_j}} $, is defined as follows:\begin{equation}
\label{minkowski}
\sum\limits_{j = 1}^K {{S_j}}  \buildrel \Delta \over = \left\{ {\sum\limits_{j = 1}^K {{s_j}} :{s_j} \in {S_j}} \right\}.
\end{equation}
That is, the sumset of the subsets $\left\{ {{S_j}} \right\}_{j = 1}^K$ is defined to be the set of all sums (using GF($q$) arithmetic) of elements taken from the subsets. When the labels $h_{ij}$ of the graph are chosen at random, the CTV message of Equation \eqref{CTV} can be considered as a sumset of random subsets of GF($q$). Noting that, $P_m$ is equivalent to the probability that a sumset of random subsets is of size $m$, when the sizes of the subsets are known. 

Finding the number of elements within the sumset as a function of ${\left\{ {\left| {{S_j}} \right|} \right\}_{j = 1}^{K}}$ is an open problem in additive combinatorics (see e.g. \cite{croot}). This stems from the structure of the field, where a symbol in a field can be obtained by multiple combinations of sums of symbols. In the following, we provide bounds on the size of the sumset.
\begin{lem}
\label{simple_bounds}
Consider $K$ non-empty subsets of GF($q$), ${\left\{ {{S_j}} \right\}_{j = 1}^K}$. Then:
\begin{equation}
\label{simp_bounds}
{\max _j}\left| {{S_j}} \right| \le \left| {\sum\limits_{j = 1}^K {{S_j}} } \right| \le \min \left( {q,\prod\limits_{j = 1}^K {\left| {{S_j}} \right|} } \right).
\end{equation}
\end{lem}
\begin{proof}
Denote by $j_0$ the index of the subset with the largest size. From the definition of a sumset in Equation \eqref{minkowski}, it is clear that there exists $a \in {\text{GF}}\left( q \right)$ such that:\[\left\{ {{s_{{j_0}}} + a:{s_{{j_0}}} \in {S_{{j_0}}}} \right\} \subseteq \sum\limits_{j = 1}^K {{S_j}}.\]
Since the elements of $S_{j_0}$ are all distinct, the lower bound follows. The upper bound is the number of sums (not necessarily distinct) within ${\sum\limits_{j = 1}^K {{S_j}} }$, which obviously bounds  $\left| {\sum\limits_{j = 1}^K {{S_j}} } \right|$ from above. 
\end{proof}
We can improve the bounds \eqref{simp_bounds}, by using the following two theorems and their corollary.
\begin{thm}
\label{cauchy_davenport}
\emph{(Cauchy-Davenport Theorem \cite{Cauchy} \cite{Davenport})} Consider the field GF($p$), $p$ prime, where $A$ and $B$ are two non-empty subsets of GF($p$). Then:
\[\left| {A + B} \right| = \left\{ {\left. {a + b} \right|a \in A,b \in B} \right\} \ge \min \left( {p,\left| A \right| + \left| B \right| - 1} \right).\]
\end{thm}

\begin{thm}
\label{karolyi}
\emph{(K\'{a}rolyi's Theorem for Finite Groups \cite{Karolyi})} Let $G$ be a finite group. $A$ and $B$ are two non-empty subsets of $G$. Denote by $p\left( G \right)$ the smallest prime factor of $\left| G \right|$. Then:
\[\left| {A + B} \right| \ge \min \left( {p\left( G \right),\left| A \right| + \left| B \right| - 1} \right).\]
\end{thm}
%The interested reader is referred to \cite{Karolyi} for a detailed proof. Using the last two theorems, we have the following corollary.

\begin{cor}
Assume a finite field GF($q$), where $q=p^s$, $p$ is prime and $s$ is a positive integer. Then:
\begin{equation}
\label{cor_bounds}
\begin{array}{l}
\max \left( {{\max _j}\left| {{S_j}} \right|,\min \left( {p,\sum\limits_{j = 1}^K {\left| {{S_j}} \right| - K + 1} } \right)} \right) \le \left| {\sum\limits_{j = 1}^K {{S_j}} } \right|\\
 \le \min \left( {q,\prod\limits_{j = 1}^K {\left| {{S_j}} \right|} } \right).
\end{array}
\end{equation}
\end{cor}
\begin{proof}
This corollary is proved by Lemma \ref{simple_bounds} and Theorems \ref{cauchy_davenport} and \ref{karolyi}, followed by induction on the number of subsets. %Note that $p=2$ for even $q$.
\end{proof}

We will denote by $B_L$ and $B_U$ the lower and upper bounds of \eqref{cor_bounds}, respectively. We have the following sufficient condition for attaining the maximal size, $q$, of the sumset.

\begin{prop}
\label{sct}
\emph{(Sufficient condition for $\left| {\sum\limits_{j = 1}^{K} {{S_j}} } \right| = q$}) If there is a pair of sets $S_a,S_b \in {\left\{ {{S_j}} \right\}_{j = 1}^{K}}$ ($a \ne b$) such that $\left| {{S_a}} \right| + \left| {{S_b}} \right| > q$, then ${\left| {\sum\limits_{j = 1}^K {{S_j}} } \right| = q}$.
\end{prop}
\begin{proof}
Consider the (Abelian) group $G = \left\{ {{\text{GF}}\left( q \right),' + '} \right\}$, i.e., $G$ consists of all $q$ elements of the field GF($q$) with the field addition operation '+'. Choose an element $g \in G$. We will now prove that there exist $s_a \in {S_a},s_b \in {S_b}$ such that $g = s_a + s_b$. Define the set $A = g - {S_b} = \left\{ { {g - {s_b}} :{s_b} \in {S_b}} \right\}$. It is clear that ${S_a} \cap A \ne \emptyset $, since $\left| {{S_a}} \right| + \left| A \right| = \left| {{S_a}} \right| + \left| {{S_b}} \right| > q$. Let $d = g-s_b$ be an element of the intersection ${S_a} \cap A$. Then:
\[d + {s_b} = \left( {g - {s_b}} \right) + {s_b} = g.\]
Now note that
\[\sum\limits_{j = 1}^K {{S_j}}  = \left\{ {{s_a} + {s_b} + \sum\limits_{i \ne a,b} {{S_i}} :{s_a} \in {S_a},{s_b} \in {S_b}} \right\},\]
and therefore  ${\left| {\sum\limits_{j = 1}^K {{S_j}} } \right| = q}$.
\end{proof}
For later use, we say that the {\it q-condition} holds if the condition of Proposition \ref{sct} is satisfied. Using the bounds $B_L$ and $B_U$ and the $q$-condition, we get the following bounds (in terms of the size of the sumset) for $P_m$:
\begin{equation}
P_m^{\left( {{\rm{max}}} \right)} = \left\{ {\begin{array}{*{20}{c}}
{\delta \left[ {m - q} \right],}&{{\text{if the } q\text{-condition holds}}}\\
{\delta \left[ {m - {B_U}} \right],}&{{\text{otherwise}}}
\end{array}} \right.
\end{equation}
\begin{equation}
P_m^{\left( {\text{min}}\right)} = \left\{ {\begin{array}{*{20}{c}}
{\delta \left[ {m - q} \right],}&{{\text{if the } q\text{-condition holds}}}\\
{\delta \left[ {m - {B_L}} \right],}&{{\text{otherwise}}}
\end{array}} \right.
\end{equation}

Using the above $P_m^{\left( {{\rm{max}}} \right)}$ resp. $P_m^{\left( {\text{min}}\right)}$ in~\eqref{Pm1} will give a lower resp. upper bound on the \textit{decoding threshold} of the QPEC, which is defined similarly to the decoding threshold of the BEC \cite{mct}.

\subsection{Equivalent formulation and formula for $Q_m$}
\label{PDVTC}

When the labels $h_{ij}$ are chosen at random, $Q_m$ is equivalent to the probability that the intersection of $d_v$ random GF($q$) subsets with sizes ${\left\{ {\left\{ {\left| {{S_j}} \right|} \right\}_{j = 1}^{{d_v} - 1},M} \right\}}$ ($M$ corresponds to the size of the set provided by the channel information) is exactly $m$. We begin with the following lemma.
\begin{lem}
\label{Im_calculation}
Assume that ${\left\{ {{S_j}} \right\}_{j = 1}^{J}}$ are subsets of a set with $q$ elements, with given sizes ${\left\{ {\left| {{S_j}} \right|} \right\}_{j = 1}^{J}}$, where $\mu  \buildrel \Delta \over = {\rm{mi}}{{\rm{n}}_j}\left| {{S_j}} \right|$. Then, the number of ways to get an intersection of size $m$ ($m=0,1,...,\mu)$ between the subsets is:
\begin{equation}
\label{Im_def}
\begin{gathered}
  {I_m}\left( {\left\{ {\left| {{S_j}} \right|} \right\}_{j = 1}^{J}};q \right) =  \hfill \\
  \sum\limits_{i = 0}^{\mu  - m} {{{\left( { - 1} \right)}^i} \cdot } \upsilon \left( {\left\{ {\left| {{S_j}} \right|} \right\}_{j = 1}^{J},m + i} \right) \cdot \ell \left( {m + i,m} \right), \hfill \\ 
\end{gathered} 
\end{equation}
where
\begin{equation}
\upsilon \left( {\left\{ {\left| {{S_j}} \right|} \right\}_{j = 1}^{J},l} \right) = \ell \left( {q,l} \right) \cdot \prod\limits_{j = 1}^{J} {\ell \left( {q - l,\left| {{S_j}} \right| - l} \right.)}. 
\end{equation}
\end{lem}

\begin{proof}
Assume a fixed set $S$ with $l$ elements taken from GF($q$). The number of ways to choose $J$ sets of sizes $\left\{ {\left| {{S_j}} \right|} \right\}_{j = 1}^{J}$ such that they all contain $S$ equals:
\begin{equation}
n\left( {\left\{ {\left| {{S_j}} \right|} \right\}_{j = 1}^{J},l} \right) = \prod\limits_{j = 1}^{J} {\ell \left( {q - l,\left| {{S_j}} \right| - l} \right)}.
\end{equation}
In addition, 
\[\begin{array}{l}
\upsilon \left( {\left\{ {\left| {{S_j}} \right|} \right\}_{j = 1}^J,l} \right) = \ell \left( {q,l} \right) \cdot \prod\limits_{j = 1}^J {\ell \left( {q - l,\left| {{S_j}} \right| - l} \right.)} \\
 = \ell \left( {q,l} \right) \cdot n\left( {\left\{ {\left| {{S_j}} \right|} \right\}_{j = 1}^J,l} \right)
\end{array}\]
is the number of combinations of subsets with sizes ${\left\{ {\left| {{S_j}} \right|} \right\}_{j = 1}^{J}}$ that have an intersection of size \textit{at least} $l$. Now, for $m=\mu$, we have $I_\mu = \ell \left( {q,\mu } \right) \cdot n\left( {\left\{ {\left| {{S_j}} \right|} \right\}_{j = 1}^{J},\mu } \right)=\upsilon \left( {\left\{ {\left| {{S_j}} \right|} \right\}_{j = 1}^{J},\mu } \right)$ ways to choose $\left\{ {{S_j}} \right\}_{j = 1}^{J}$ such that their intersection is of size $\mu$. For $m=\mu-1$, we have
\[\begin{array}{l}
{I_{\mu  - 1}}\left( {\left\{ {\left| {{S_j}} \right|} \right\}_{j = 1}^J} \right)\\
 = \upsilon \left( {\left\{ {\left| {{S_j}} \right|} \right\}_{j = 1}^J,\mu  - 1} \right) - \ell \left( {\mu ,\mu  - 1} \right) \cdot \upsilon \left( {\left\{ {\left| {{S_j}} \right|} \right\}_{j = 1}^J,\mu } \right)
\end{array}\]
ways to choose $\left\{ {{S_j}} \right\}_{j = 1}^{J}$ such that their intersection is of size $\mu-1$. This was obtained by subtracting intersections of size $\mu$ from sets with intersection of size at least $\mu-1$. Continuing in the same fashion (essentially, we use the inclusion-exclusion principle), we get:
\begin{equation}
\begin{array}{l}
{I_{\mu  - t}}\left( {\left\{ {\left| {{S_j}} \right|} \right\}_{j = 1}^J} \right)\\
 = \sum\limits_{i = 0}^t {{{\left( { - 1} \right)}^i} \cdot } \upsilon \left( {\mu  - t + i} \right) \cdot \ell \left( {\mu  - t + i,\mu  - t} \right),
\end{array}
\label{Imu}
\end{equation}
for $t=0,1,...,\mu$. Index shifting leads to the desired result.

\end{proof}
%\end{lemma}

We are now ready to provide an exact formula for $Q_m$.

\begin{thm}
Assume that ${\left\{ {{S_j}} \right\}_{j = 1}^{J}}$ are subsets of GF($q$) with given sizes ${\left\{ {\left| {{S_j}} \right|} \right\}_{j = 1}^{J}}$. Further assume w.l.o.g. that each subset contains the symbol $0$ (as can be assumed due to the all-zero codeword assumption). Then, the probability for an intersection of size $m$ ($m=1,2,...,\mu={\text{min}}(\left\{ {\left| {{S_j}} \right|} \right\}_{j = 1}^{J}$) between the subsets is:
\begin{equation}
\label{Qm_def}
\begin{gathered}
  {Q_m}\left( {\left\{ {\left| {{S_j}} \right|} \right\}_{j = 1}^{J};q} \right) =  \hfill \\
  \left\{ {\begin{array}{*{20}{c}}
  {\frac{{{I_{m - 1}}\left( {\left\{ {\left| {{S_j}} \right| - 1} \right\}_{j = 1}^{J};q - 1} \right)}}{{\sum\limits_{l = 1}^\mu  {{I_{l - 1}}\left( {\left\{ {\left| {{S_j}} \right| - 1} \right\}_{j = 1}^{J};q - 1} \right)} }},}&{{\text{if}}\hspace{7pt} \mu  > 1} \\ 
  {\delta \left[ {m - 1} \right],}&{{\text{otherwise}}} 
\end{array}} \right. \hfill \\ 
\end{gathered} 
\end{equation}
%{Q_m}\left( {\left\{ {\left| {{S_j}} \right|} \right\}_{j = 1}^{n + 1};q} \right) = \left\{ {\begin{array}{*{20}{c}}
%{\frac{{{I_{m - 1}}\left( {\left\{ {\left| {{S_j}} \right| - 1} \right\}_{j = 1}^{n + 1};q - 1} \right)}}{{\sum\limits_{m = 1}^\mu  {{I_{m - 1}}\left( {\left\{ {\left| {{S_j}} \right| - 1} \right\}_{j = 1}^{n + 1};q - 1} \right)} }},}&{{\text{if}}\hspace{7pt} \mu  > 1}\\
%{\delta \left[ {m - 1} \right],}&{{\text{otherwise}}}
%\end{array}} \right.

where $I_m$ is defined in Equation \eqref{Im_def}.  %means the number of ways to choose $J$ subsets with sizes ${\left\{ {\left| {{S_j}} \right| - 1} \right\}_{j = 1}^{J}}$ from a set with $q-1$ elements such that the subsets intersection size equals $m-1$.
\end{thm}
\begin{proof}
This is a result of Lemma \ref{Im_calculation}. We shift $m$ to $m-1$  since the symbol $0$ belongs to each set. Moreover, instead of $q$ possible elements to choose from, we have only $q-1$ possible ones, since $0$ is already taken.
\end{proof}

\section{Approximation models for $P_m$}
\label{pm_approx}

So far, we provided an exact formula for $Q_m$ and bounds for $P_m$. An exact expression for $P_m$ is likely difficult to find. In this section, we discuss appropriate models for approximating ${P_m}\left( {\left\{ {\left| {{S_j}} \right|} \right\}_{j = 1}^K} \right)$ from Equation \eqref{Pm1} (where $K=d_c-1$). We begin with a simple \textit{balls-and-bins model}, and later refine it with a tighter model we term as the \textit{union model}.

In the balls and bins model \cite{Mitz}, there is a set of balls and a set of bins. Each bin is assumed to be picked independently and uniformly at random for each ball. In our case, there are $N = \prod\limits_{j = 1}^{K} {\left| {{S_j}} \right|}$ sums within the sumset \eqref{minkowski}, each leading to an element in GF($q$). Using this model, we approximate $P_m$ as the probability that $N$ balls are assigned to exactly $m$ ($m=1,2,...,q$) bins. This probability can be calculated in a recursive manner. 

\begin{figure}[h]
\begin{center}
\includegraphics[scale=0.44]{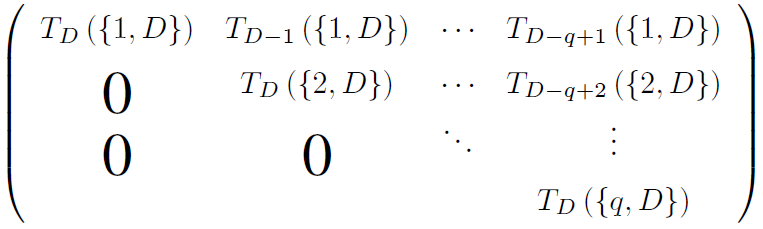}
\caption{The matrix $\bf{\Gamma}(D)$}
\label{TD_matrix}
\end{center}
\end{figure}

In the following, we provide a formulation of this model as a \textit{Markov process}, leading to an easy calculation of the desired probabilities for the model. First, define:
\begin{equation}
\label{Tm_def}
{T_m}\left( {\left\{ {\left| {{S_j}} \right|} \right\}_{j = 1}^K} \right) = {\rm{ }}\frac{{{I_m}\left( {\left\{ {\left| {{S_j}} \right|} \right\}_{j = 1}^K} \right)}}{{\sum\limits_{l = 0}^{\mu } {{I_l}\left( {\left\{ {\left| {{S_j}} \right|} \right\}_{j = 1}^K} \right)} }}
\end{equation}
using $I_m$ from Equation \eqref{Im_def} ($q$ was omitted for convenience), where $\mu  \buildrel \Delta \over = {\rm{mi}}{{\rm{n}}_j}\left| {{S_j}} \right|$ . $T_m$ is the probability that the intersection of randomly chosen subsets $\left\{ {{S_j}} \right\}_{j = 1}^K$ of a set of size $q$ is of size $m$. Using $T_m$, we define the matrix ${\bf{\Gamma}}={\bf{\Gamma}}(D)$, with its elements being ${\left( {\bf{\Gamma }} \right)_{i,j}} = {T_{i + D - j}}\left( {\left\{ {i,D} \right\}} \right)$. ${T_{i + D - j}}\left( {\left\{ {i,D} \right\}} \right)$ is defined to be zero when $i + D - j < 0$ or $i + D - j > q$. ${\bf{\Gamma}}$ is an upper triangular matrix, as can be seen in Figure \ref{TD_matrix}. This matrix is a stochastic matrix describing a {\it Markov chain} with $q$ {\it states} (in fact, ${\bf{\Gamma }}$ is an \textit{absorbing} Markov chain) 

Consider ${{\bf{\Gamma }}_A} = {\bf{\Gamma }}\left( 1 \right)$. ${{{\mathbf{\Gamma }}_A}}$ contains non-zero elements on indices of the form $(i,i),(i,i + 1)$ only, where ${\left( {{{\mathbf{\Gamma }}_A}} \right)_{i,i}} = \frac{i}{q}$ and ${\left( {{{\mathbf{\Gamma }}_A}} \right)_{i,i + 1}} = 1 - \frac{i}{q}$. The $q$ {\it states} of the chain defined by ${{\bf{\Gamma }}_A}$ correspond to the number of occupied bins. Now, define the probability vector ${{\bf{g}}^{\left( l \right)}} = \left( {g_1^{\left(l \right)},g_2^{\left(l \right)},...,g_q^{\left( l \right)}} \right)$ over the states of ${{\bf{\Gamma }}_A}$, where ${{\mathbf{g}}^{\left( 1 \right)}} = \left( {1,0,...,0} \right)$. We have the following relation:
\begin{equation}
\label{g_calc}
{{\mathbf{g}}^{\left( l \right)}} ={{\mathbf{g}}^{\left( 1 \right)}} {\mathbf{\Gamma }}_A^{l - 1}
\end{equation}
where in this case ${{\bf{g}}^{\left( l \right)}}$ is simply the first row of ${{\bf{\Gamma }}_A^{l-1}}$. ${{g}}_m^{\left( N \right)}$ is the probability that $m$ bins are occupied in the balls and bins model, given  that the number of balls is $N$.

Taking into account the lower bound $B_L$ and the {\it q-condition}, we get the following approximation for $P_m$:
\begin{equation}
\label{pm_balls}
P_m^{\left( {{\rm{balls}}} \right)} = \left\{ {\begin{array}{*{20}{c}}
{\delta \left[ {m - q} \right],}&{{\text{if the } q\text{-condition holds}}}\\
{\frac{{g_m^{\left( N \right)}}}{{\sum\limits_{i = {B_L}}^q {g_i^{\left( N \right)}} }},}&{{\text{otherwise}}}
\end{array}} \right.
\end{equation}

%We are now ready to calculate the expected number of occupied bins according to $P_m^{\left( {{\rm{balls}}} \right)}$. Define the vector $\boldsymbol{\eta }$, which has $q$ elements, such that its first $B_L-1$ elements are zero, and the remaining elements are ones. The expected number of occupied bins according to $P_m^{\left( {\text{balls}} \right)}$ is:
%
%\begin{equation}
%{E^{\left( {{\text{balls}}} \right)}} = \left\{ {\begin{array}{*{20}{c}}
%  {q,}&{{\text{if } q\text{-condition holds}}} \\ 
%  {\frac{{{\boldsymbol{\eta }} \cdot {{\mathbf{g}}^{\left( N \right)}}}}{{\sum\limits_{j = {B_L}}^q {g_j^{\left( N \right)}} }},}&{{\text{otherwise}}} 
%\end{array}} \right.
%\end{equation}
%where ${{{\mathbf{g}}^{\left( N \right)}}}$ can be calculated efficiently using Equation \eqref{g_calc}.
%\begin{figure*}%
%\centering
%\begin{subfigure}{.4\columnwidth}
%\includegraphics[width=\columnwidth]{allmodelsq4}%
%\caption{cap a}%
%\label{subfiga}%
%\end{subfigure}\hfill%
%\begin{subfigure}{.4\columnwidth}
%\includegraphics[width=\columnwidth]{allmodelsq5}%
%\caption{cap b}%
%\label{subfigb}%
%\end{subfigure}\hfill%
%\end{figure*}
According to the balls and bins model, each ball (sum) is assigned to a bin independently. However, it is clear that the sums that appear within the sumset \eqref{minkowski} can be divided into $N/{\max _j}\left| {{S_j}} \right|$ sets of ${\max _j}\left| {{S_j}} \right|$ \textit{distinct} elements (since a fixed partial sum from the other sets is translated by ${\max _j}\left| {{S_j}} \right|$ distinct elements). To take this into account, we next define the \textit{union model}.

In the union model, a \textit{set} of ${\max _j}\left| {{S_j}} \right|$ balls is assigned to a \textit{set} of ${\max _j}\left| {{S_j}} \right|$ bins. Similarly to the balls and bins model, the union model can also be formulated in terms of a Markov process, using the matrix ${{\bf{\Gamma }}_B} = {\bf{\Gamma }}\left( {\max _j}\left| {{S_j}} \right| \right)$. Define the probability vector ${{\bf{u}}^{\left( l \right)}} = \left( {u_1^{\left( l \right)},u_2^{\left( l\right)},...,u_q^{\left( l \right)}} \right)$ over the states of ${{\bf{\Gamma }}_B}$, where ${{\mathbf{u}}^{\left( 1 \right)}}$ has $1$ in position ${\max _j}\left| {{S_j}} \right|$ and its remaining elements are zeros. We have the following relation:
\begin{equation}
{{\mathbf{u}}^{\left( l \right)}} = {{\mathbf{u}}^{\left( 1 \right)}}{\mathbf{\Gamma }}_B^{l - 1}
\end{equation}
where in this case ${{\bf{u}}^{\left( l \right)}}$ is simply the ${\left( {{{\max }_j}\left| {{S_j}} \right|} \right)^{{\rm{th}}}}$ row of ${{\bf{\Gamma }}_B^{l-1}}$. ${u_m^{\left( l \right)}}$ is the probability that $m$ bins are occupied after $l$ sets of size ${\max _j}\left| {{S_j}} \right|$ were assigned (at random and independently, set-by-set) to the bins. As in the balls and bins model, we use the lower bound $B_L$ and the {\it q-condition}, to get the following approximation for $P_m$:
\begin{equation}
P_m^{\left( {{\rm{union}}} \right)} = \left\{ {\begin{array}{*{20}{c}}
{\delta \left[ {m - q} \right],}&{{\text{if the } q\text{-condition holds}}}\\
{\frac{{u_m^{\left( N/{\max _j}\left| {{S_j}} \right|\right)}}}{{\sum\limits_{i = {B_L}}^q {u_i^{\left( N/ {\max _j}\left| {{S_j}} \right|\right)}} }},}&{{\text{otherwise}}}
\end{array}} \right.
\end{equation}

As in the case of BEC/QEC, we have a threshold phenomenon \cite{mct} for the QPEC. The operational meaning of the threshold can be thought of as the maximal allowed fraction of partially known (up to $M$ levels) symbols in a $q$-level flash memory, such that all the symbols will be decoded correctly after a sufficient number of iterations.

\begin{figure}
        \centering
        \begin{subfigure}[b]{0.4\textwidth}
                \includegraphics[width=\textwidth]{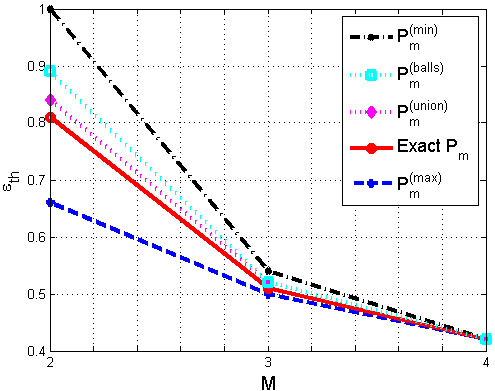}
                \caption{$q=4$}
               \label{q4results}
        \end{subfigure}%
        ~ %add desired spacing between images, e. g. ~, \quad, \qquad etc.

        \begin{subfigure}[b]{0.4\textwidth}
                \includegraphics[width=\textwidth]{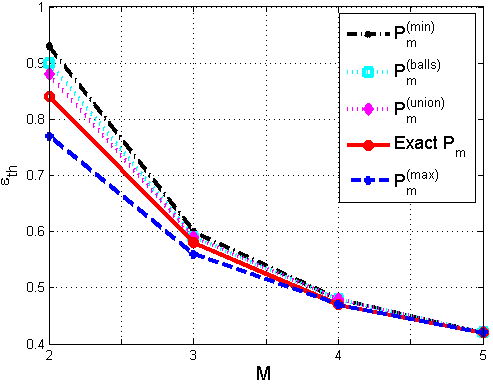}
                \caption{$q=5$}
                \label{q4results}
        \end{subfigure}

        \caption{The threshold $\varepsilon_{\rm th}$ for $(3,6)$ LDPC code}
        \label{fig_results}
\end{figure}

In Figure \ref{fig_results}, we provide the threshold (denoted $\varepsilon_{\text {th}}$) for a regular $(3,6)$ LDPC code, calculated using the density evolution equations \eqref{Pm1} and \eqref{Qm_DE} for $q=4$ and $q=5$. The exact $P_m$ was calculated numerically (by averaging over all possible assignments of sets and counting the number of distinct elements in the corresponding sumset). The lower bound corresponds to $P_m^{\left( {{\rm{max}}} \right)}$ and the upper bound corresponds to $P_m^{\left( {{\rm{min}}} \right)}$.

As can be seen, both the balls and bins model and the union model appear to serve as an upper bound for $\varepsilon_{th}$, with the union model being tighter. As mentioned in Section \ref{de_analysis}, the same threshold is obtained for all the models when $M=q$.

% \begin{figure}[h]
%\begin{center}
%\includegraphics[scale=0.62]{allmodelsq3.png}
%\caption{$q=3, \varepsilon_{th}$ comparison (same result for all models)}
%\end{center}
%\end{figure}
%
%\begin{figure}[h]
%\begin{center}
%\includegraphics[scale=0.62]{allmodelsq4.png}
%\caption{$q=4, \varepsilon_{th}$ comparison}
%\end{center}
%\end{figure}
%
%\begin{figure}[h]
%\begin{center}
%\includegraphics[scale=0.62]{allmodelsq5.png}
%\caption{$q=5, \varepsilon_{th}$ comparison}
%\end{center}
%\end{figure}
%
%\begin{figure}[h]
%\begin{center}
%\includegraphics[scale=0.62]{allmodelsq7.png}
%\caption{$q=7, \varepsilon_{th}$ comparison}
%\end{center}
%\end{figure}
%%
%\begin{figure}[h]
%\begin{center}
%\includegraphics[scale=0.62]{allmodelsq8.png}
%\caption{$q=8, \varepsilon_{th}$ comparison}
%\end{center}
%\end{figure}
%
%\begin{figure}[h]
%\begin{center}
%\includegraphics[scale=0.62]{allmodelsq4.png}
%\caption{The threshold $\varepsilon_{\rm th}$ for $q=4$, $(3,6)$ LDPC code}
%
%\vspace{-15pt}
%\label{fig_results}
%\end{center}
%\end{figure}

%\begin{figure}
%        \centering
%        \begin{subfigure}{0.33\textwidth}
%                \includegraphics[width=\textwidth]{allmodelsq4}
%                \caption{$q=4$}
%                \label{fig:gull}
%        \end{subfigure}%        
%        ~ %add desired spacing between images, e. g. ~, \quad, \qquad etc.
%        %  (or a blank line to force the subfigure onto a new line)
%        
%        \begin{subfigure}{0.33\textwidth}
%                \includegraphics[width=\textwidth]{allmodelsq5}
%                \caption{$q=5$}
%                \label{fig:tiger}
%        \end{subfigure}
%
%     \caption{The threshold $\varepsilon_{\text{th}}$ for a regular ($3,6$) LDPC code \label{fig_results}
%}
%\end{figure}

\section{Conclusion}
\label{conc}
In this paper, we defined a new channel - the QPEC - motivated by multilevel NVMs. We provided an appropriate belief propagation decoder for this channel when used with LDPC codes, with the corresponding density evolution equations. We developed approximation models for these equations, since their exact analysis is closely related to an open problem in additive combinatorics. The results show the importance of these models, which appear to provide a good approximation.
%Suggestions for further research include:
%
%\begin{enumerate}
%\item {{Additional models for $P_m$.}}
%\item { {Capacity-achieving degree distributions for QPEC with $M<q$.}}
%\item { Analysis of a {\textit{Constrained} QPEC, in which the possible values in case of erasure are drawn according to some non-uniform distribution (e.g., known to be consecutive).}}
%\end{enumerate}

% or


\bibliographystyle{IEEEtran}
	\bibliography{Cohen_Cassuto_QPEC}

\end{document}